\documentstyle[twocolumn,prl,aps]{revtex}

\begin{document}

\draft

\twocolumn[\hsize\textwidth\columnwidth\hsize\csname
@twocolumnfalse\endcsname

\title{Packing Transitions in Nanosized Li Clusters}

\author{Ming-Wen Sung}
\address{Department of Physics, University of California,
San Diego, La Jolla, CA 92093}

\author{Ryoichi Kawai}
\address{Department of Physics, University of Alabama at Birmingham,
Birmingham, AL 35294}

\author{John H. Weare}
\address{Department of Chemistry, University of California,
San Diego, La Jolla CA, 92093}

\date{\today}

\maketitle

\begin{abstract}
Packing transitions in the lowest energy structures
of Li clusters as a function of size
have been identified via simulated annealing.
For $N$ $\leq$ 21, the large $p$ character of Li leads to
unexpected ionic structures.
At $N$ $\approx$ 25, a packing pattern based on
interpenetrating 13-atom icosahedra
and similar to that of Na and K appears.
This pattern persists until at $N$=55, where another transition
to a structure based on a Mackay icosahedron occurs.
For clusters of size 55 and 147, the optimized FCC structure representative
of the bulk is still slightly higher in energy than the optimal MIC.
\end{abstract}


]
\narrowtext
The change in cluster properties such as structure as size increases
is a problem of
considerable interest\cite{Jena92}.
Recently, the abundance spectra of alkali clusters
have been measured\cite{Martin90,Bowen92} up to very large sizes.
The jellium model\cite{deHeer87} for sizes $N<1500$ and
structural stabilities based on icosahedral or cuboctahedral (FCC)
packing
for large clusters ($1500<N<22000$)
were successfully used to explain the observed intensity maxima.
However, these measurements give no direct structural information.
Electron microscope measurements of atomic positions have been
reported for gold clusters for sizes above 400 atoms\cite{Iijima86}.
Cuboctahedral and icosahedral geometries were observed supporting
high density packing for large metal clusters.
In the bulk limit, Li and Na undergo a
martensitic transformation from
BCC to a close-packed crystal at low temperatures\cite{Smith87,Schwarz90}.

On the other hand, {\it ab initio} calculations of geometric and electronic
properties of Li clusters within the Hartree-Fock (HF)
\cite{Sugino90},
the SCF-CI method\cite{Boustani87,Koutecky91} and the generalized valence bond
methods\cite{McAdon85} and spectroscopic measurements\cite{Dugourd91}
find that small Li clusters have open structures rather than dense packing.
Similar results have
been found for other simple metals\cite{Koutecky91,Andreoni91,Jones91}.

In this article we report optimal structures for Li clusters
from the small to large particle regime.
An unbiased search, simulated annealing, was used to obtain lowest energy
configurations.
The Car-Parrinello molecular dynamics simulation which combines
dynamics and the local density approximation\cite{Car85}
has been successfully used
in cluster calculations\cite{Andreoni91,Kawai90,Andreoni94}.
However, the artificial heating of the electron wavefunction makes
application of this method to metal clusters difficult.
The calculations reported here
use an {\it ab initio} molecular dynamics (AIMD) simulation
based on
a conjugate gradient algorithm\cite{Stich89,Kresse93} to update the
electronic structure following each step of atomic motion.
With this method we can take much larger time steps,
limited only by the time scale of classical motion of
the atoms.
Except for this modification, the method
is the same as in previous AIMD simulations\cite{Andreoni91,Kawai90,Kawai91}.

The calculations were performed within the pseudopotential
local spin density (LSD) method\cite{Kawai91,Hamann89}.
A simple cubic supercell with lattice constant $a$ = 30.0 -- 45.0 a.u.
and a cutoff energy of 11.2 Ry provided adequate convergence.
Predictions for structure, energies and vibrational
frequencies for Li$_2$ and Li$_3$ were
close to experimental data\cite{Kawai94}, e.g., for Li$_2$, D$_e$=1.03
eV (LSD) and 1.07 eV(Exp.);
$\omega$ = 341 cm$^{-1}$ (LSD) and 351 cm$^{-1}$ (Exp.).
Annealing was started from a randomly generated initial geometry and
the system was allowed to equilibrate at a temperature about twice the bulk
melting temperature (700--800 K) for about 10 picoseconds.
After this initial equilibration the cluster was cooled as slowly as
possible ($10^{14}$ -- $10^{15}$ K/sec) until solid-like dynamics appeared.
At this point the system was quenched to the nearest minimum energy structure.
For clusters of sizes less than 55, repeated annealing
led to the same or very similar degenerate structures.
Above this size selected geometries
were optimized by steepest descent or conjugate gradient minimization.

The electronic shell structures and the large HOMO-LUMO gap at
the magic numbers $N$=8, 20 and 40 are in good agreement with the jellium
model as in the Na and Be systems\cite{Andreoni91,Kawai90}.
As in prior alkali metal calculations, the orbital energies
are fairly insensitive to the detailed atomic positions\cite{Andreoni91}.

However, Li and the other first row elements
have large $sp$ interactions due to the lack of
$p$ orbitals in the core\cite{Jones89}.
The pseudopotential for the Li 2$p$ state is deeper and less repulsive at
the origin than for other group I elements.
This results in more participation of $p$ orbitals in bonding.
For example, the $sp$ promotion for Li clusters
is much larger and increases more quickly with size than in Na and K.

For the Li system, the increased $sp$ promotion results in
optimal structures with increased polarization
and charge transfer between interior and surface atoms and
very different
geometries when compared with the Na and K systems.
For example, at the LSD level, a three dimensional structure has
the lowest energy at $N$=5 \cite{Kawai94} for Li, whereas heavier alkali
elements remain planar till $N$=6\cite{Andreoni91}.

More remarkably, the enhanced charge transfer in Li$_8$
leads to the unusual centered trigonal prism (CTP)
structure with a capping atom illustrated in Fig.~\ref{fig_Li8}a.
This structure is very different from the dicapped octahedral
(dodecahedron) structure (Fig.~\ref{fig_Li8}b) previously found
for Na \cite{Andreoni91} and also for K\cite{Sung94}.
The central atom is negatively charged by nearly $-e$
(the charge is estimated by population analysis\cite{Kawai91}).  The
extra electron on the central atom occupies $p$ type orbitals.
The balance of positive charge is distributed almost equally among
surface atoms.  Because of this charge distribution, the surface atoms
are attracted to the center atom, whereas the attraction between
the surface atoms is reduced.
The structure is similar to that of ionic solvation and
resembles that of the ionic molecule TaF$_7^{2-}$
\cite{Dekock}.
It has not been identified in other calculations\cite{Koutecky91}.

The charge density in the Li$_8$ CTP structure is largest
near the interior atom in
contrast to the high charge density between surface atoms in the
dicapped octahedral structure of Na and K.
In order for such a structure to be stable, the pseudopotential
of the central atom must be able to accommodate the
increased density and the nodal property of the cluster orbitals
at the central atom.
Since the additional charge on the central atom occupies
atomic $p$ orbitals, the $p$ component of the pseudopotential
is responsible for the difference between Li and other group I elements.
In the Li system the $sp$ interaction
is enhanced by the deep minimum in the $p$ component of the
pseudopotential, whereas the relatively shallow $p$ potential
of the other group I elements suppresses the $sp$ promotion.
Therefore, for the Na and K systems,  the CTP structure
is not stable and spontaneously distorts to a dicapped triangular prism
\cite{Sung94} which has no central atom and very
little charge transfer.
When the $p$ component of the Li pseudopotential is
artificially made more repulsive, the CTP structure
distorts in a way similar to Na and K.

For the Li systems with large charge
transfer and short bonds (because of the relatively small core sizes of the
first row elements) there is also stability from the
Coulomb interactions.
While the geometric structures of the Na and Li systems are qualitatively
different, the nodal structures of the cluster orbitals and the overall
cluster distortions for both systems
are well described by the jellium model\cite{deHeer87}, which
focuses on the kinetic energy and shape of the uniform positive
background charge.

The CTP structure of Fig.~1a is remarkable not only
for its ionic character but also because it initiates a pattern of growth
which continues with particle addition.
In Figs.~\ref{fig_SS}a-\ref{fig_SS}d we
illustrate this pattern schematically.
Fig. \ref{fig_SS}a is a representation of the CTP structure
(Fig.~\ref{fig_Li8}a) of Li$_8$.
Li$_{9}$ forms a structure similar to Li$_{8}$ with one more atom in the
``solvation shell'' (Fig.~\ref{fig_SS}b).
The eight surface atoms surrounding the central atom form two squares in
staggered positions as in
TaF$_8^{3-}$\cite{Dekock}.
All these molecules
minimize the repulsion between charged surface atoms.
At $N$=14 (Fig.~\ref{fig_SS}c), a second internal
atom appears with face sharing CTP structures.
The Li$_{13}$ structure is similar to that of Li$_{14}$
with one capping atom
removed.   The 13-atom icosahedral structure (IC13) is
0.035 eV/atom higher in energy.
The local coordination around the
two internal atoms in Li$_{11-13}$\cite{Kawai92} is identical to that of
Li$_{14}$.

At $N$=16, another atom is in the first shell, increasing the
coordination around the interior atoms by one.
Above this size, the third solvation shell is gradually developed and
a fully solvated structure is produced at $N$=19 (Fig.~\ref{fig_SS}d).
However, while the total charge transfer per atom
is relatively constant,
as the system size increases the localization of the charge
on the second shell decreases, decreasing the Coulombic stabilization
per atom due to charge separation and thus the relative stability
of the CTP pattern.

Between $N$=22 and 25, simulated annealing produces a new packing pattern.
The highest coordination number jumps to 12, indicating a close-packed
structure is established.  This dense packing is based on
interpenetrating 13-atom local icosahedra (LIC) formed by
an interior atom and its 12 nearest neighbors.
These structures may be constructed from an IC13 core
by placing additional surface atoms
either on the top of vertex atoms (12 sites)
or above the center of triangular faces (20 sites)
(Fig.~\ref{fig_LIC}~{LIC}).
Structures based on CTP packing were optimized as possible candidates
for the optimal structure for clusters larger than 20.  These structures
distort to higher coordinated capped pentagon structures with somewhat
higher energies than the optimal structures.
At this point the structures for the  Li, Na and K
systems are all similarly based on the LIC packing\cite{Sung94}.
For the Na and K systems, the CTP structure was never stable so the LIC
structure begins at a lower particle number\cite{Andreoni91,Sung94}.

In order to make a quantitative structural characterization of the minimum
energy clusters, a Blaisten-Farojas (BF) analysis\cite{Jonsson88} was
carried out.  In this analysis the local geometry around a pair of atoms
is characterized by four indices.  If the atoms are within a certain distance
($R_{c}$ = 6.0 -- 6.5 $a_{0}$), an index 1 is assigned.
Otherwise, the first index is 0.  Next, the number of the nearest neighbors
that the two atoms have in common is counted.  This  number is assigned as
the second index.  The third index specifies the number of bonds
between the nearest neighbors.  A fourth index is added to provide a unique
correspondence in case different geometries lead to the same indices.
For example, choosing the capping atoms in the pentagonal bipyramid that is an
important element of the icosahedral structure, the BF index is 1551.
The BF index for the nearest neighbors of FCC crystal is 1421.

The relative abundance of these indices as a function of cluster size is
illustrated in Fig.~\ref{fig_BF}.
The index 1551 is very rare for the size N$<25$,
showing the lack of pentagonal structures.  It suddenly
becomes abundant at $N$=25 due to the appearance of LIC packing.
For $N$=55, it decreases due to the appearance of a new structure based on
Mackay icosahedron (MIC, discussed in more detail below).
On the other hand, the index 1421 is prevalent for $N$$<$20 but
completely disappears between $N$=25 and 45.
It appears again at $N$=55 due to the appearance of MIC which has a
packing similar to that of FCC.

The LIC packing pattern continues at least up to Li$_{45}$ at which
point all the triangular faces and vertex faces may be occupied,
producing a highly symmetrical $I_d$ structure.
For the systems $N$=25, 30 and 40, perfect LIC structures were obtained
directly from annealing calculations.
However, simulated annealing of Li$_{45}$ converged to a slightly
disordered structure which was slightly more stable than the perfect LIC.
For $N$=45 and above, there are significant problems with
LIC packing which lead
to a new packing paradigm.
In the perfect Li$_{45}$ structure,
the distance between surface atoms in the optimized
perfect structure is large
($\approx$ 6.0 $a_{0}$) relative to the bulk ($\approx$ 5.88 $a_0$).
More importantly, the center to first shell and the intershell
bonds have become very short ($\approx$ 5.1 $a_{0}$ and 5.3 $a_{0}$).
The effect of this strain on the system is apparent in the spherically
averaged density which for Li$_{45}$ is much larger in the
center of the cluster than the bulk value.
The low surface atomic density of the LIC pattern produces high local
energy density in the center.  However, if bond lengths
in the surface are to be kept near those of the bulk, compression
of the interior bonds must occur.
At some point this compression
will become so unfavorable that a new packing pattern will be optimal.

A packing that eases these difficulties and that can extend to larger sizes
can be achieved by placing atoms on the 30 bridge sites of the
IC13 core cluster instead of on the triangle faces,
as illustrated in  Fig.~\ref{fig_LIC}~{MIC}.
Occupying all these sites on a IC13 core will form a 55-atom Mackay
icosahedron (MIC) (13 core + 12 vertex + 30 bridge atoms).

For $N$=45, an incomplete MIC structure is somewhat higher in energy than
the perfect LIC structure ($\Delta$E = .022 eV/atom).
However, for $N$=55 the perfect MIC structure is lower in energy
than the optimized LIC structures we obtained from simulated annealing.

Another possible structure for the 55 atom system is
a fragment of an FCC crystal terminated mostly
by (111) facets (a cuboctahedron).
We found that the ``perfect'' FCC structure
is higher in energy than the ``perfect'' MIC structure and
distorts with a small activation barrier to the optimized
MIC structure.  Presumably this is because of the larger number of nearest
neighbor interactions in the MIC pattern; 234 for MIC vs. 216 for FCC.

Similar structural transitions have been predicted for Lennard-Jones (LJ)
clusters by Northby\cite{Northby87}.  However for these systems, the transition
from
face capping to bridging (LIC to MIC)
takes place at a smaller particle
number, $N$=31.
In order to provide additional support for the stability of the LIC structure
for Li above $N$=31 (preference for face capping), we calculated the
energy of Li$_{25}$, Li$_{30}$,
Li$_{40}$, and Li$_{45}$ quenched from the Northby structures.
In all cases the highly ordered LIC patterns found from simulated annealing
gave lower energies.

As the system grows further, our results suggest that the
MIC structure forms a core and further growth proceeds as atoms are added
to the surface in either bridged sites or face capping sites.
However, calculations of Li$_{65}$ show that growth immediately after
$N$=55 is based on face capping.
At $N$=147 both MIC and FCC have ``perfect'' (geometrically spherical)
structures.
However, for 147 the MIC has a slightly lower energy
($\Delta$E = .022 eV/atom).
There is a suggestion in the electron density
and the interior bond lengths (5.38 au for MIC vs. 5.72 au for FCC) that a
transition to FCC is imminent.
Calculations show that
such a transition results in a decrease in the internal electron density.
The near degeneracy of the MIC and FCC structures is consistent with the
observation of both structures in electron microscope studies
for silver clusters\cite{Iijima86}.

This work was partly supported by Office of Naval Research
through NOO14-91-J-1835.  The CRAY Y-MP at Naval Oceanographic Center
and the CM-5 at Naval Research Laboratory
were used for computations.


\begin{thebibliography}{10}

\bibitem{Jena92}
  {\em Physics and Chemistry of Finite Systems: From Clusters to Crystals},
  edited by R. Jena, S.~N. Khanna, and B.~K. Rao (Kluwer Academic, The
  Netherlands, 1992).

\bibitem{Martin90}
H. G{\"{o}}hlich, T. Lange, T. Bergmann, and T.~P. Martin, Phys. Rev. Lett.
  {\bf 65},  748  (1990).

\bibitem{Bowen92}
C. Brechignac and {\it et al}, in Ref.\cite{Jena92}, p.369.

\bibitem{deHeer87}
W.~A. de~Heer, W.~D. Knight, M.~Y. Chou, and M.~L. Cohen,  in {\em Solid State
  Physics}, edited by F. Seitz and D. Turnbull (Academic, New York, 1987),
  Vol.~40, p.\ 93.

\bibitem{Iijima86}
S. Iijima and T. Ichihashi, Phys. Rev. Lett. {\bf 56},  616  (1986).

\bibitem{Smith87}
H.~G. Smith, Phys. Rev. Lett. {\bf 58},  1228  (1987).

\bibitem{Schwarz90}
W. Schwarz and O. Blaschko, Phys. Rev. Lett. {\bf 65},  3144  (1990).

\bibitem{Sugino90}
O. Sugino and H. Kamimura, Phys. Rev. Lett. {\bf 65},  2696  (1990).

\bibitem{Boustani87}
I. Boustani and {\it et al}, Phys. Rev. B {\bf 35},  9437  (1987).

\bibitem{Koutecky91}
V. Bona{\u{c}}i{\'{c}}-Kouteck{\'{y}}, P. Fantucci, and J. Kouteck{\'{y}},
  Chem. Rev. {\bf 91},  1035  (1991).

\bibitem{McAdon85}
M.~H. McAdon and W.~A. Goddard, Phys. Rev. Lett. {\bf 55},  2563  (1985).

\bibitem{Dugourd91}
P. et~al, Phys. Rev. Lett. {\bf 67},  2638  (1991).

\bibitem{Andreoni91}
U. R{\"{o}}thlisberger and W. Andreoni, J. Chem. Phys. {\bf 94},  8129  (1991).

\bibitem{Jones91}
R.~O. Jones, Phys. Rev. Lett. {\bf 67},  224  (1991).

\bibitem{Car85}
R. Car and M. Parrinello, Phys. Rev. Lett. {\bf 55},  2471  (1985).

\bibitem{Kawai90}
R. Kawai and J.~H. Weare, Phys. Rev. Lett. {\bf 64},  81  (1990).

\bibitem{Andreoni94}
U. Rothlisberger, W. Andreoni, and M. Parrinello, Phys. Rev. Lett. {\bf 72},
  665  (1994).

\bibitem{Stich89}
I. {\v{S}}tich, R. Car, M. Parrinello, and S. Baroni, Phys. Rev. B {\bf 39},
  4997  (1989).

\bibitem{Kresse93}
G. Kresse and J. Hafner, Phys. Rev. B {\bf 47},  558  (1993).

\bibitem{Kawai91}
R. Kawai and J.~H. Weare, J. Chem. Phys. {\bf 95},  1151  (1991).

\bibitem{Hamann89}
D.~R. Hamann, Phys. Rev. {\bf B40},  2980  (1989).

\bibitem{Kawai94}
R. Kawai, J.~F. Tombrello, and J.~H. Weare, Phys. Rev. A {\bf 49},  4236
  (1994).

\bibitem{Jones89}
R.~O. Jones and O. Gunnarson, Rev. Mod. Phys. {\bf 61},  689  (1989).

\bibitem{Sung94}
M. Sung, R. Kawai, and J.~H. Weare, to be submitted.

\bibitem{Dekock}
R.~L. DeKock and H.~B. Gray, {\em Chemical Structure and Bonding} (University
  Science Books, Mill Valley, CA, 1989), p.\ 491.

\bibitem{Kawai92}
R. Kawai, M.-W. Sung, and J.~H. Weare,  in Ref.\cite{Jena92}, p.\ 441.

\bibitem{Jonsson88}
H. J{\"{o}}nsson and H.~C. Anderson, Phys. Rev. Lett. {\bf 60},  2295  (1988).

\bibitem{Northby87}
J.~A. Northby, J. Chem. Phys. {\bf 87},  6166  (1987).

\end{thebibliography}

\begin{figure}
\caption{The lowest energy structures of (a) Li$_8$ and (b) Na$_8$
(a dodecahedron).}
\label{fig_Li8}
\end{figure}

\begin{figure}
\caption{Schematic representations of (a) Li$_8$, (b) Li$_9$, (c)
Li$_{14}$, and (d) Li$_{19}$.  Shaded circles indicate vertex atoms
of trigonal prisms and solid circles are center atoms in the
prisms.  Open circles represent atoms capping the sides of the prisms.
The shaded
and open circles form a solvation shell around the solid circle.}

\label{fig_SS}
\end{figure}

\begin{figure}
\caption{Blaisten-Farojas analysis.  The upper and lower curves
indicate the abundance of pentagonal bipyramids and rectangular
bipyramids in the lowest energy structure, respectively.
$\triangle$ and $\Box$ indicate
the LIC at N=55 and the bulk FCC, respectively. }
\label{fig_BF}
\end{figure}

\begin{figure}
\caption{(a) 45-atom LIC structure.  LIC axes passing through the center
of IC13 have been drawn.  (b) Successive MIC models with complete outer
layers are constructed by growing each of the 20 tetrahedra constituting
the primitive IC13.}
\label{fig_LIC}
\end{figure}

\end{document}